\documentclass[letter]{aa}  
\usepackage{natbib,twoopt}
\usepackage[breaklinks=true]{hyperref} 
\bibpunct{(}{)}{;}{a}{}{,}             
\makeatletter
  \newcommandtwoopt{\citeads}[3][][]{\href{http://adsabs.harvard.edu/abs/#3}%
    {\def\hyper@linkstart##1##2{}%
     \let\hyper@linkend\@empty\citealp[#1][#2]{#3}}}
  \newcommandtwoopt{\citepads}[3][][]{\href{http://adsabs.harvard.edu/abs/#3}%
    {\def\hyper@linkstart##1##2{}%
     \let\hyper@linkend\@empty\citep[#1][#2]{#3}}}
  \newcommandtwoopt{\citetads}[3][][]{\href{http://adsabs.harvard.edu/abs/#3}%
    {\def\hyper@linkstart##1##2{}%
     \let\hyper@linkend\@empty\citet[#1][#2]{#3}}}
  \newcommandtwoopt{\citeyearads}[3][][]%
    {\href{http://adsabs.harvard.edu/abs/#3}
    {\def\hyper@linkstart##1##2{}%
     \let\hyper@linkend\@empty\citeyear[#1][#2]{#3}}}
\makeatother
\usepackage{graphicx}
\usepackage{epstopdf}
\usepackage{txfonts}
\begin{document}

   \title{Temperature constraints on the coldest brown dwarf known: WISE 0855-0714}

   \author{J.~C.~ Beam\'in
          \inst{1,2,3}
          \and
          V.~ D.~ Ivanov \inst{2}
          \and
          A.~ Bayo\inst{4,5}
          \and 
          K.~Mu\v{z}i\'c\inst{2}
          \and 
          H.~M.~J.~ Boffin \inst{2}
          \and
          F. Allard\inst{6}
          \and
          D. Homeier\inst{6}
          \and 
          D.~ Minniti \inst{1,3,7}
          \and 
          M.~ Gromadzki \inst{3,5}
          \and
          R.~ Kurtev\inst{5,3}
          \and
          N.~ Lodieu \inst{8,9}
          \and
          E.~L.~Martin \inst{10}
          \and 
          R.~A.~Mendez \inst{2,11}
          }
\institute{
Instituto de Astrof\'isica, Facultad de F\'isica, Pontificia Universidad 
Cat\'olica de Chile, Casilla 306, Santiago 22, Chile
\email{jcbeamin@astro.puc.cl}
\and
European Southern Observatory, Ave. Alonso de Cordoba 3107, Casilla 19001, Santiago, Chile
\and
Millennium Institute of Astrophysics, Santiago, Chile
\and
Max Planck Institut f\"{u}r Astronomie, K\"{o}nigstuhl 17, 69117, Heidelberg, Germany
\and
Instituto de F\'isica y Astronom\'ia, Universidad de Valpara\'iso, Av. Gran Breta\~{n}a 1111, Playa Ancha, Casilla 5030 Valpara\'iso, Chile
\and
Centre de Recherche Astrophysique de Lyon, UMR 5574, CNRS, Universit\'e de Lyon, \'Ecole Normale Sup\'erieure de Lyon, 46 All\'ee d'Italie, F-69364 Lyon Cedex 07
\and
Departamento de Ciencias Fisicas, Universidad Andres Bello, Republica 220, Santiago, Chile
\and
Instituto de Astrof\'isica de Canarias (IAC), Calle V\'ia L\'actea s/n, E-38200 La Laguna, Tenerife, Spain
\and
Departamento de Astrof\'isica, Universidad de La Laguna (ULL), E-38206 La Laguna, Tenerife, Spain
\and
Centro de Astrobiología (INTA-CSIC), Carretera de Ajalvir km 4, 28550 Torrej\'on de Ardoz, Madrid, Spain
\and
Universidad de Chile, Departamento de Astronom\'ia, Casilla 36-D, Santiago, Chile
}

   \date{Received June xx, 2014; accepted October xx, 2014}

 
\abstract
{Nearby isolated planetary mass objects are beginning to be discovered, but
their individual properties are poorly constrained because their low surface 
temperatures and strong molecular self-absorption make them extremely faint.}
{We aimed to detect the near infrared emission of the coldest brown dwarf 
(BD) found so far, WISE0855$-$0714, located $\sim$2.2 pc away, and to improve 
its temperature estimate (T$_{\rm eff}$= 225-260 K) from a comparison with 
state-of-the-art models of BD atmospheres.}
{We observed the field containing WISE0855-0714 with HAWK-I at the VLT in the $Y$ band.  
For BDs with T$_{\rm eff}<$500\,K theoretical models predict strong signal (or rather 
less molecular absorption) in this band.}
{WISE0855-0714 was  not detected in our Y-band images, thus placing an upper limit on its 
brightness to  Y>24.4 mag at 3-$\sigma$ level, leading to Y-[4.5]>10.5.
Combining this limit with previous detections and upper limits at other wavelengths, 
 WISE0855$-$0714 is confirmed as the reddest BD detected, further supporting its status
 as the coldest known brown dwarf.
We applied spectral energy distribution fitting with collections of models from two independent groups
for extremely cool BD atmospheres  leading to an  effective temperature of T$_{\rm eff}<$250\,K,
}
{}


\maketitle
%

\section{Introduction}

Over the last two decades we witnessed the discovery of the first brown 
dwarfs  \citep[BDs; e.g.,][]{Stauffer1994,Rebolo1995,Nakajima1995,Basri1996}, and a fast
development of BD science, from extending the definition of stellar/substellar 
spectral classes to L-T-Y 
\citep{Kirkpatrick1999,Martin1999,Burgasser2006,Cushing2011}, to the studies
of climatic variations in their atmospheres \citep[e.g.,][]{Morales2006,Artigau2009,Radigan2014}. 
Within the last year we also saw how Wolf 359 (CN Leonis) was downgraded  from 
the third to the fifth closest system to the Sun. It was surpassed by the BD binary  
Luhman 16AB \citep{Luhman2013} and by \object{WISE J085510.83–071442.5} 
(hereafter WISE0855), the coldest known BD to date  \citep{Luhman2014b}.

These and other recent discoveries 
\citep{Scholz2014,Luhman2014c,Kirkpatrick2014,Cushing2014,Perez2014} suggest that the 
census of nearby very low mass stars and BDs is still incomplete, even within 
a few parsecs from the Sun. The study of these ultra-cool dwarfs in the solar 
neighborhood and in young clusters have helped to better understand and 
constrain the stellar-substellar initial mass function 
\citep[see][and references therein]{Luhman2012}, the 
evolution of their physical properties \citep{Burrows2001}, and the nature of 
their atmospheres \citep{Biller2013}.

The proximity of these new objects helps to characterize them better. For 
example, high quality optical spectra and optical polarimetry can be obtained 
despite the extremely red colors, to estimate their T$_{\rm eff}$, to measure the 
radial velocities, and to constrain the presence of scattering disks 
\citep[e.g.,][]{Kniazev2013}. The higher level of brightness with respect to
more distant objects makes it possible to demonstrate the spotty surface of
these objects; \citet{Biller2013} and \citet{Burgasser2014} showed that the variability of Luhman 16B
is related to spots and $\sim$30\% cloud coverage; \citet{Crossfield2014} 
traced the clouds moving on the surface of Luhman 16B with high resolution 
spectroscopy and Doppler imaging. Accurate astrometry is also feasible:
\citet{Boffin2014} reported indications of a planetary mass companion in the 
same system. Recently, \citet{Faherty2014}  confirmed the strong lithium absorption
observed by \citet{Luhman2013} in the primary, and also detected it in the secondary 
component of the system (the first detection of lithium in a T dwarf)
constraining the age of the system to 0.1$-$3.0 Gyr, revealing more details 
in the atmospheric and cloud properties of the system. 

\citet{Luhman2014b} discovered WISE0855 through its high proper motion (PM), 8.1$\pm$0.1 
\arcsec yr$^{-1}$, by comparing multi-epoch  Wide-field Infrared Survey Explorer
\cite[{\it WISE;}][]{Wright2010} observations.  The 4.5$\mu$m absolute magnitude
 indicated an effective temperature in the range T$_{\rm eff} \sim$225$-$260 K, 
making it extremely faint even in the near-infrared (NIR) wavelengths, 
with a $J$-band upper limit of $\sim$23 mag. Combined {\it WISE} and {\it Spitzer Space 
Telescope} \citep{Werner2004} observations yielded a parallax of 0.454$\pm$0.045 
\arcsec (2.20$^{+0.24}_{-0.20}$ pc). \citet{Wright2014} published new NEOWISE-R 
(NEO WISE Reactivation mission) observations, increasing the baseline, and 
improving the parallax (0.448$\pm$0.033 \arcsec or 2.23$^{+0.17}_{-0.15}$ pc) 
and the PM (8.09$\pm$0.05 \arcsec yr$^{-1}$). They also reported 
$H$-band imaging non-detection with a limit of 22.7 mag.

 Given the few positional measurements used to derive PM and parallax, and to refine
 its temperature estimate, we attempted to detect WISE0855 
in the $Y$ band. This particular band was selected based on the theoretical ultra-cool BD atmosphere 
models of \citet{Morley2014} that predicted weaker molecular absorption in objects 
with T$_{\rm eff}$ $<$300 K at $\lambda$ $\sim$1 $\mu$m (see their Fig. 12) making them 
brighter in $Y$ than in $J$. In Section 2 we describe our observations. In Section 3 we place an upper limit on 
the T$_{\rm eff}$ from spectral energy distribution 
(SED) fitting. The last section summarizes our results.

\section{Observations}

WISE0855 was imaged with the High Acuity Wide-field K-band Imager 
\citep[HAWK-I; ][]{Pirard2004,Kissler2008} on Unit Telescope 4 of ESO’s Very 
Large Telescope. This camera has four 2048x2048 HAWAII-2RG arrays and a plate 
scale of $\sim$0.106 \arcsec pix$^{-1}$. The target was placed on chip number 2 (as this chip shows
less structure).
The observations were collected on the nights of May 10 (18 images), 17 (36 
images), and 28 (18 images), 2014. Each image was the average of four frames with 
integration of 29.2 seconds, giving a total exposure time of 140.16 minutes. 

\begin{figure}
\centering
\includegraphics[scale=0.3]{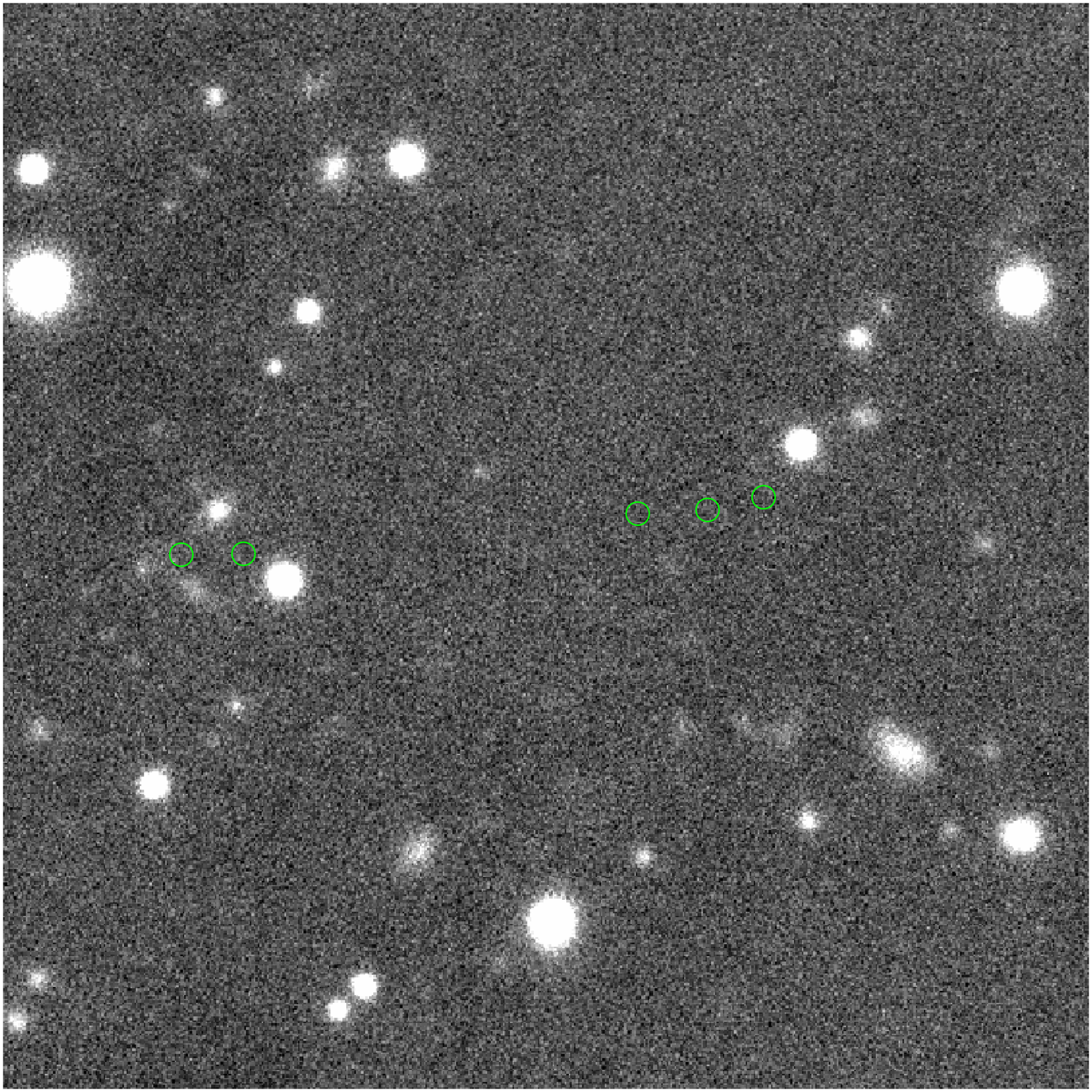}
\caption{A 1$\arcmin$x1$\arcmin$ field of our final $Y$-band image around the 
expected position of WISE0855 (north is up, east is left). Green circles mark the positions of the source 
given by \citet{Luhman2014b} and \citet{Wright2014} from the {\it WISE} and {\it Spitzer} 
observations. The  first circle on the right shows the position obtained with the WISE satellite on May 5, 2014.
Our observations were made between  May 10 and  May 26, 2014, and the source would have moved less 
than 10 pixels in this image and hence would be visible at the right edge of  that circle.
The circle sizes are larger than the uncertainties in position.}
\label{fig:image}%
\end{figure}

Reduction of individual frames was performed using the HAWK-I pipeline. 
This includes dark subtraction, division by the flat
field, and sky subtraction. The files from each night were treated separately,
with calibrations taken on the next day, or during the next evening twilight in the case of flats.
The sky subtraction was performed in the
double-pass scheme, where the first background estimate is
subtracted from the science frames, which are then combined and used
to create an object mask. The mask is used on the second pass to exclude
the objects in the field during the creation of the final sky.
Finally, the astrometric calibration of individual reduced frames 
was refined using bright 2MASS stars in the field as references.
Two attempts of co-adding the images with 
SWarp v2.38.0 software \citep{Bertin2002} were performed.
First, we co-added them, leaving the reference stars fixed, to obtain a reliable detection limit on non-moving
sources. The second attempt accounted for the PM of WISE0855: the images were taken with a 
difference in time of 17 days, implying a motion of the BD of $\sim$0.4\arcsec.  
We co-added the images fixing the pixels of the expected position of the moving source to obtain 
the photometric limit on the source of interest.  The FWHM for the combined image is 0.6\arcsec ($\sim$6\,pixels).
A zoom-in from this image, of the region around the WISE0855 location, is shown in 
Fig. \ref{fig:image}\footnote{Figure produced with Aladin \citep{Bonnarel2000}}.

UKIRT standards were observed each night at similar airmass as WISE0855: 
\object{FS 16} on May 10, \object{FS 136} and \object{FS 19} on May 17, and 
\object{FS 19} and \object{FS 132} on May 28. They were reduced in the same way 
as the science data, and the fluxes of the standard stars were measured with 
large apertures (typically $\sim$ 5 \arcsec) to avoid having to apply aperture 
corrections. The final photometric zero-point derived from these images was  26.333 $\pm$ 0.037 mag.

\begin{figure}
\centering
\includegraphics[height=6.5cm,keepaspectratio]{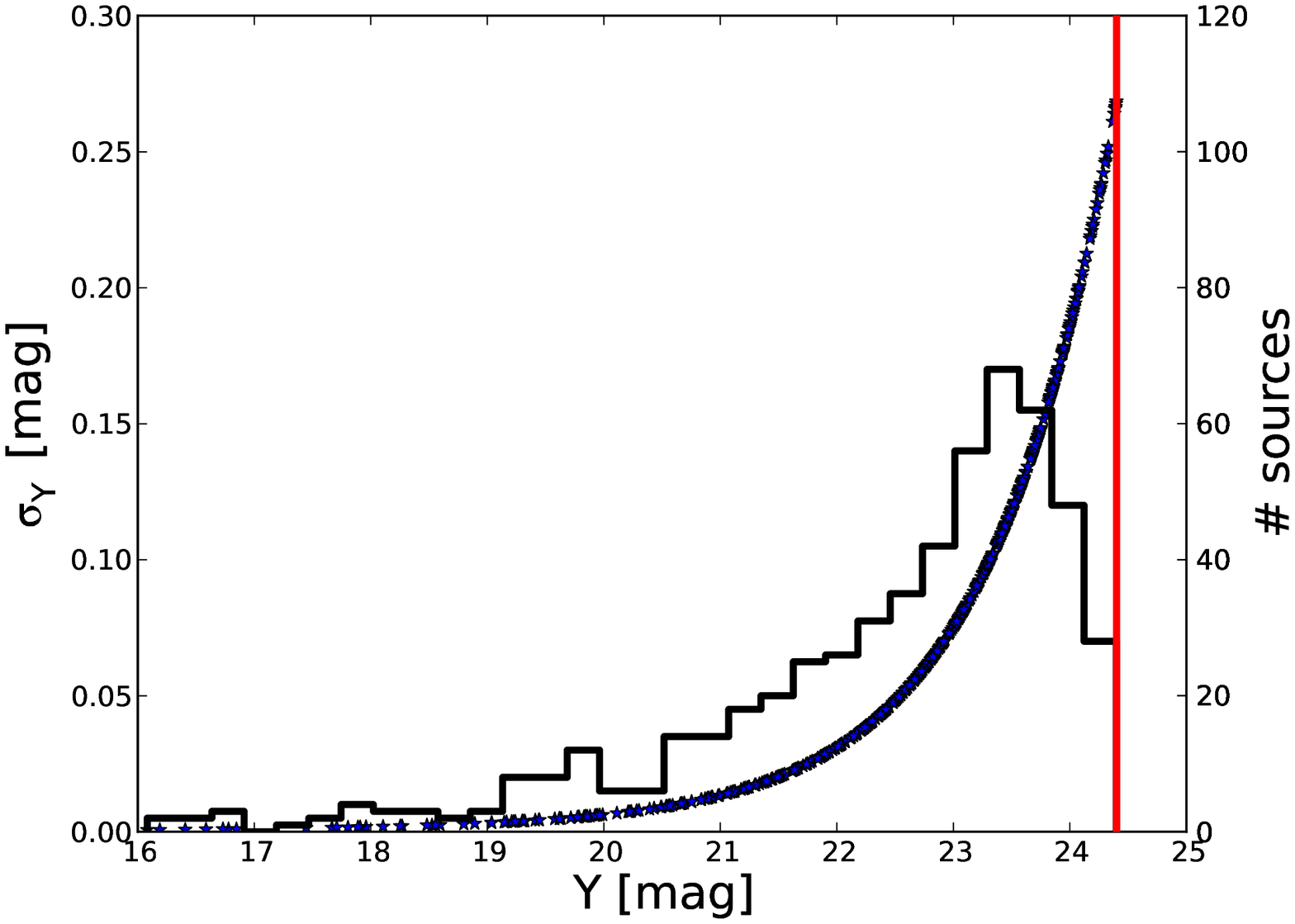}
\caption{Luminosity function of the sources in the field surrounding WISE0855 (black
solid line) and errors on individual sources (blue stars). The red vertical 
line shows our detection limit at the location of the source. The photometric 
errors at the faint limiting magnitude are $\sim$0.25 -- 0.3\,mag. }
\label{fig:LF}%
\end{figure}

We performed aperture photometry with SExtractor ver. 2.19.5 \citep{Bertin1996} to both final
co-added images with relaxed constraints on the detection (only five pixels above threshold and
 3-$\sigma$ above threshold detection, activating the algorithm to clean spurious detections).
The luminosity function of the field containing WISE0855 is shown in Fig. 
\ref{fig:LF}.
The error in the photometry as a function of the magnitude is plotted with star symbols
and the vertical red line at  $Y$= 24.4 shows the magnitude upper limit for WISE0855 at a 3-$\sigma$ level.
This limit was determined as three times the standard deviation of the sky background level around the 
expected position. 
We adopt this as our final upper limit in the forthcoming analysis.
All sources detected by the software (fainter than  $Y>$24.4 mag.) were removed for clarity. 

\section{Temperature estimates}

\begin{figure}
\centering
\includegraphics[scale=0.45]{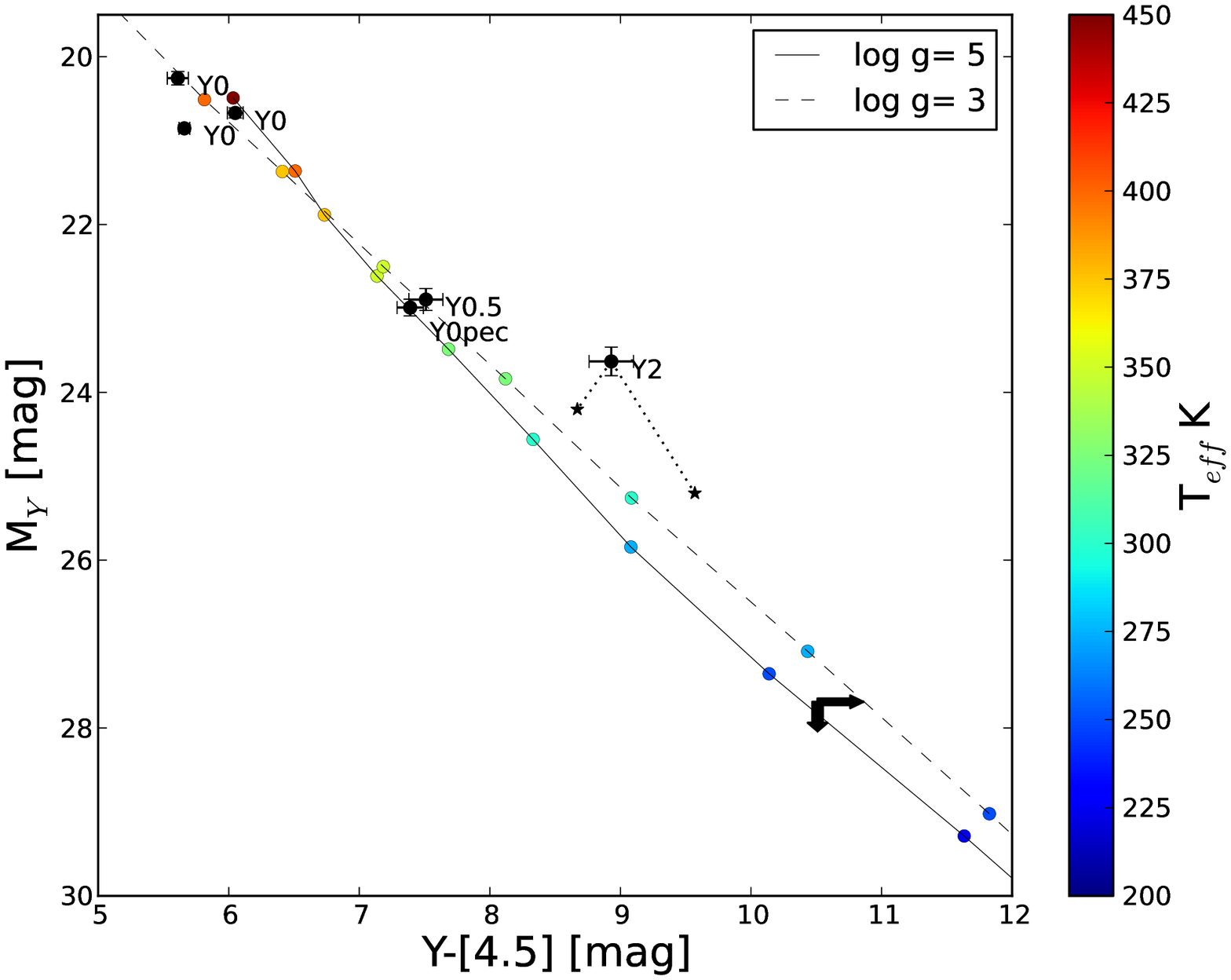}
\caption{ Absolute color magnitude diagram of Y dwarfs with available measurements in the Y-band filter from \citet{Leggett2013}. 
A color shift was not applied to account for differences in the pass-band from Gemini-NIRI data to  VLT-HAWK-I data;
distances were taken from \citet{Leggett2013} and \citet{Beichman2014}.
The arrow at the bottom right shows the upper limit presented in this work, making WISE0855 the reddest BD.
 The stars show the the individual components of the tentative Y2 binary (\object{WISEPCJ1828+2650}), proposed to 
explain the over-luminosity in the MIR.
Colored points shows two of the models from \citet{Morley2014}, with 50$\%$ cloud coverage, a sedimentation factor  
f$_{sed}$= 5, and log g= 3 and 5, connected with black dashed and solid line for clarity.
}
\label{fig:cmd}%
\end{figure}

\begin{figure}
\centering
\includegraphics[scale=0.6]{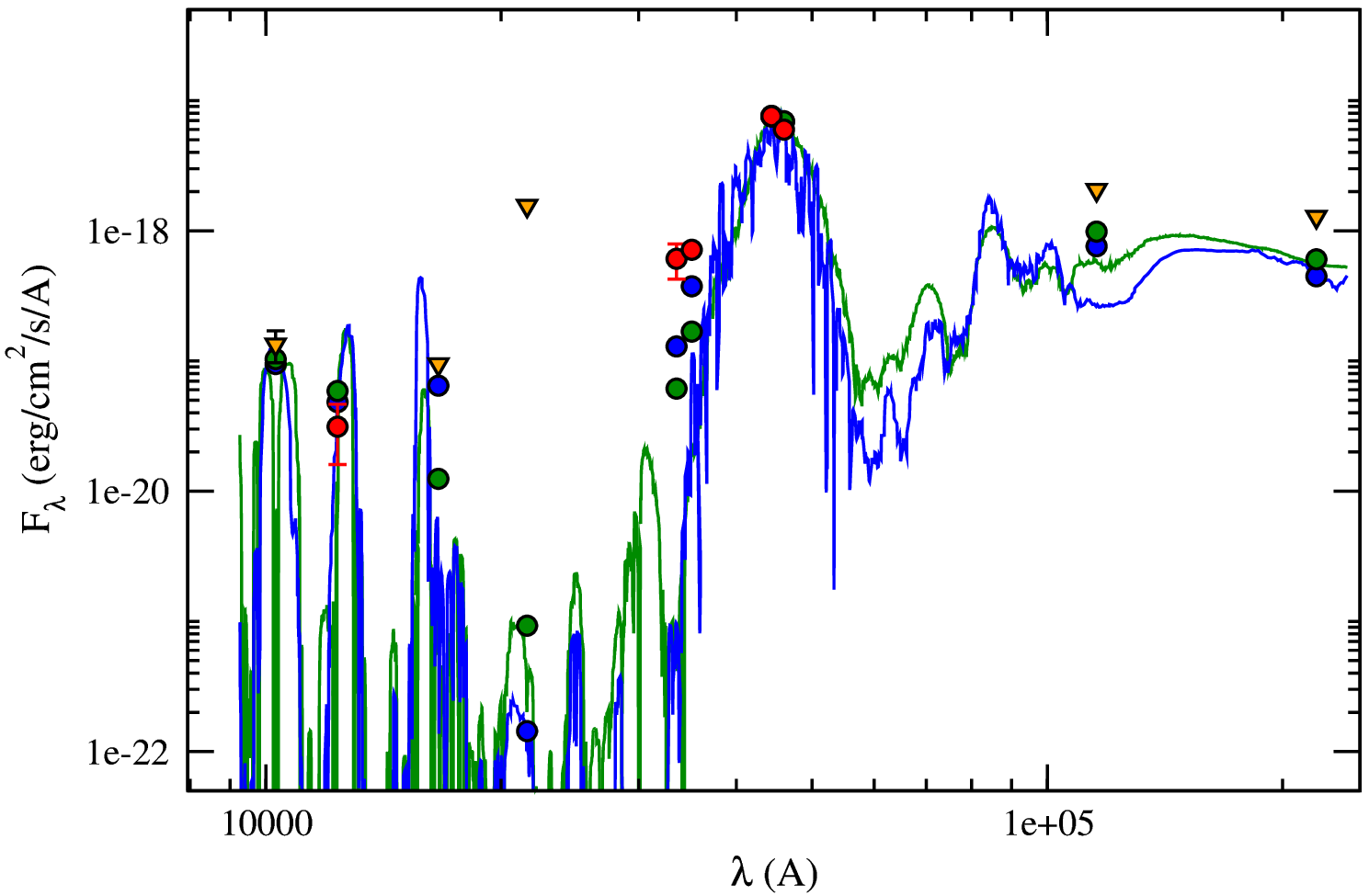}
\caption{Spectral energy distribution of WISE0855. Yellow triangles show the magnitude upper limits in $Y$ (this work) 
 and $H$, $K\rm _S$ , W3, and W4 \citep{Luhman2014b, Wright2014}. Red circles show measurements from J band \citep{Faherty2014b}; WISE bands W1,W2 and
{\it Spitzer} IRAC channels 1 and 2. Blue and green points represent the magnitudes derived from convolving the synthetic spectrum with
the corresponding filter transmission curves for each model. Finally, the blue curve is the best fitting BT-settl model with temperature 
T$_{\rm eff}$= 240~K, log g=4, and radius R= 1.17 R$_{Jup}$, conforming to all upper limits, and the green curve is the best fitting \citet{Morley2014}
model, with  T$_{\rm eff}$= 250 K, log g= 4, f$_{sed}$= 7, and 50$\%$ of cloud coverage (see \citealt{Morley2014} for more details in the parameters).
See Sec.\,3 for detailed description of the models.
}
\label{fig:sed}%
\end{figure}

\begin{table}
\caption{Photometric measurements and upper limits for WISE0855.}             
\label{Tab:mags}      
\centering                          
\begin{tabular}{@{ }c@{}c@{ }c@{}c@{ }c@{ }}        
\hline\hline                 
 Band      &  Magnitude & S/N & Flux  & Ref.\\    
           &            &     & ~~[erg s$^{-1}$ cm$^{-2}$ $\AA{}^{-1}$]~~ & \\     
\hline                        
            $Y$ & >24.4  &   <3               & 1.012E-19 & (1)\\
            $J$ & 25.   &   2.6               & 3.129E-20 & (2) \\
            $H$ & >22.7  &    <3              & 9.423 E-20 & (4)\\
            $K\rm_S$ & >18.6 &   <3           & 1.555 E-18 & (3)\\
            W1  & ~~17.819$\pm$0.327~~ & $\gtrsim$3     & 6.096 E-19 & (4)\\
            W2  & 14.016$\pm$0.048  & $\gtrsim$20     & 5.977 E-18 & (4)\\
            W3  & $\geq$11.25 &    <5         & 2.060 E-18 & (3)\\
            W4  & >9      &        <5        & 1.278 E-18 & (3)\\
            IRAC [3.6] & 17.44$\pm$0.05& $\gtrsim$20  & 7.139 E-19 & (3)\\
            IRAC [4.5] & 13.89$\pm$0.02 & $\gtrsim$50 & 7.579 E-18 & (3)\\
\hline                                   
\end{tabular}
\tablebib{
(1)~This work; (2) \citet{Faherty2014b} ; (3) \citet{Luhman2014b}; (4) \citet{Wright2014}}
\end{table}

We compare our $Y$-band upper limit of WISE0855 with the 
$Y$-band photometry of Y dwarfs from \citet{Leggett2013}\footnote{The
HAWK-I $Y$-band filter is shifted blueward by 0.011$\mu m$ compared 
to the UKIDSS system, and it is 0.004$\mu m$ bluer compared to the 
NIRI $Y$-band filter used in the \citet{Leggett2013} study, we decided
not to transform the magnitudes to UKIDSS system and to compare directly 
between HAWK-I $Y$-band and  NIRI $Y$-band as the filter difference between 
HAWK-I $Y$ and UKIDSS would generate larger uncertainties. To do that, 
we added 0.17 mag to the values listed in table 2 of \citet{Leggett2013}} on
a M$_Y$ vs. $Y$-[4.5] CMD ( Fig. \ref{fig:cmd};  [4.5] measurements come from 
{\it Spitzer}). This object is $\gtrsim$1 mag redder than the Y2 type 
 \object{WISEPCJ1828+2650}, the reddest/coldest object before WISE0855 
was discovered. Measurements for colder dwarfs are not 
available yet, so our result only indicates that WISE0855 
belongs to a later type than Y2.

Only a handful of theoretical models are publicly available in the
range of extremely low temperatures, inhabited by WISE0855.
The field was pioneered by the now outdated AMES-COND model grid of \citep{Allard2001}.
As an example, it yields  a temperature T$_{\rm eff} \sim$100 K and a raius of $\sim$10\,R$_{\rm Jup}$ 
(which is too large) for WISE0855.

For this study we have used an updated version of the 2014 BT-Settl grid in preparation, which adds
opacities of additional condensates forming at lower temperatures including Na$_2$S, Cr, ZnS, and KCl,
and extends the cloud model calculation to temperatures where first water ice and then ammonium
hydrosulphide (NH$_4$SH) form. In contrast to the models of \citep{Morley2014}, which incorporate
a similar set of condensates, but model the emergent spectrum as a mixture of cloudy and cloud-free
patches, these models assume a homogeneous cloud deck with a thickness controlled by the balance
of upmixing, settling, and depletion in the same formalism as used by the \citep{Allard2012, Allard2013}  models.
In particular, we tested a mini-grid of updated BT-Settl models with temperatures from 380\,K down to 200\,K and surface gravities
in the range $\log(g)$= 3.5$-$4.5.  These models were fed into the VOSA \citep{Bayo2008} database
to compute the synthetic photometry for all filters with available  photometry or upper-limits (see Table \ref{Tab:mags}).
The upper limits were incorporated by adding a penalizing factor for
their violation, so only fits that met all the constraints were considered.
The dilution factor M$_{\rm d}=$ (R/D)$^2$, where R is the object radius and D is the distance to the object, was 
optimized to minimize the residuals in the brute-force fit to the detections; the radius obtained for the best model was 1.17 R$_{Jup}$.
 We also fitted the new models from \citet{Morley2014}\footnote{models available at \url{http://www.ucolick.org/~cmorley/cmorley/Models.html}}.
  and almost the same upper limit in temperature was found; the best fit was obtained using the following parameters: 
T$_{\rm eff}$= 250 K, $\log(g)$= 4, f$_{sed}$= 7 (efficiency of sedimentation). 

The T$_{\rm eff}$= 240~K, log g= 4, BT-settl model (see Fig. \ref{fig:sed}) yields the best $\chi^2 \sim$8, although still
far from an ideal fit. Lower temperatures also lead to acceptable fits, suggesting that 
the temperatures can easily be lower. However, it is highly unlikely that WISE0855 is warmer; none of 
the models with T$_{\rm eff}$>250\,K produces an acceptable fit because of the
stringent $Y$- and $J$-band constraints (see Fig. \ref{fig:cmd}).

\section{Discussion and conclusions}

We observed WISE0855 in $Y$ band with HAWK-I at the ESO VLT,
integrating for $\sim$2.33 hours. The source was not detected, adding to the string of upper limits
in $J$, $H$, $K\rm _S$, and  W3 bands (Table \ref{Tab:mags}; and most recently in z-band; \citealt{Kopytova2014}).
We placed a  3-$\sigma$ limit of  $Y$= 24.4 mag, making WISE0855 the 
object with the reddest NIR to MIR colors up to date. A SED 
fit with state-of-the-art models of extremely cool BD atmospheres was applied to all available measurements 
and upper limits: T$_{\rm eff}\sim$250 K is the maximum 
temperature allowed by any of the models we tested.

We investigated whether WISE0855 might belong to any of the known young moving groups: 
we calculated the Galactic UVW velocities\footnote{Galactic UVW Calculator, created by David Rodriguez:
\url{http://www.astro.ucla.edu/~drodrigu/UVWCalc.html}} for radial velocities in the range 
from V$_{rad}$= $-$200 to 200\,km\,s$^{-1}$, adopting the PM and the 
distance from \citet{Luhman2014b} and \citet{Wright2014}. There is no V$_{rad}$ for which 
the UVW would match those of any of the moving groups listed in \citet{Torres2008}.

WISE0855 belongs to a new, but possibly numerous \citep{Wright2014} group of nearby extremely cool BDs 
and free-floating planetary mass objects. Ongoing and future deep multi-epoch surveys such as VISTA ESO public surveys  
(VISTA Hemisphere survey -VHS, and VISTA Variables in the V\'ia L\'actea -VVV)
UKIDSS, Pan-STARSS, Gaia, LSST, and Euclid, among others, might help to find new members of the solar neighborhood 
\citep{Burningham2011,Aller2013,Beamin2013,Bruijne2014,Smith2014}.

\begin{acknowledgements}
Funded by Project IC120009 "Millennium Institute of Astrophysics (MAS)" of 
Iniciativa Cient\'ifica Milenio del Ministerio de Econom\'ia, Fomento y Turismo 
de Chile. J.C.B., D.M., R.K., acknowledges support from: PhD Fellowship from CONICYT, 
Project FONDECYT No. 1130196, and  grant 1130140 respectively. 
M.G. is financed by the GEMINI-CONICYT Fund, allocated to Project 32110014.
NL was funded by the Ram\'on y Cajal fellowship number 08-303-01-02 and
his research is supported by project AYA2010-191367 from the Spanish Ministry of Economics and
Competitiveness (MINECO).
R. A. Mendez acknowledges partial support from project PFB-06 CATA-CONICYT.
D.H. acknowledges support from the European Research Council under the European Community’s
Seventh Framework Programme (FP7/2007-2013 Grant Agreement no. 247060).
Model atmosphere computations were performed at the {P\^ole Scientifique de Mod\'elisation Num\'erique} (PSMN) of
the {\'Ecole Normale Sup\'erieure de Lyon} and at the {Gesellschaft f{\"u}r Wissenschaftliche Datenverarbeitung
G{\"o}ttingen} in collaboration with the Institut f{\"u}r Astrophysik G{\"o}ttingen.
This work is based on Director’s Discretionary observations made with the
European Southern Observatory Telescopes at the Paranal Observatory under 
programme 293.C-5011(A). 
This publication makes use of data products from the Two Micron All Sky Survey, which is a joint project of the University 
of Massachusetts and the Infrared Processing and Analysis Center/California Institute of Technology, funded by 
NASA and NSF.
This publication makes use of VOSA, developed under the Spanish Virtual Observatory project supported from the Spanish 
MICINN through grants AYA2008-02156 and AYA2011-24052.
\end{acknowledgements}



\begin{thebibliography}{}
\bibitem[Allard et al. (2001)]{Allard2001} Allard, F., Hauschildt, P.~H., Alexander, D.~R., Tamanai, A., \& Schweitzer, A.,\ 2001, \apj, 556, 357
\bibitem[Allard et al.(2012)]{Allard2012} Allard, F., Homeier, D., Freytag, B., \& Sharp, C.~M.\ 2012, EAS Publications Series, 57, 3 
\bibitem[Allard et al.(2013)]{Allard2013} Allard, F., Homeier, D., Freytag, B., et al.\ 2013, Memorie della Societa Astronomica Italiana 
Supplementi, 24, 128 
\bibitem[Aller et al.(2013)]{Aller2013} Aller, K.~M., Kraus, A.~L., Liu, M.~C., et al.\ 2013, \apj, 773, 63
\bibitem[Artigau et al.(2009)]{Artigau2009} Artigau, {\'E}., Bouchard, S., Doyon, R., \& Lafreni{\`e}re, D.\ 2009, \apj, 701, 1534
\bibitem[Basri et al.(1996)]{Basri1996} Basri, G., Marcy, G.~W., \& Graham, J.~R.\ 1996, \apj, 458, 600 
\bibitem[Bayo et al. (2008)]{Bayo2008} Bayo, A., Rodrigo, C., Barrado Y Navascu{\'e}s, D., et al.\ 2008, \aap, 492, 277
\bibitem[Beam{\'{\i}}n et al.(2013)]{Beamin2013} Beam{\'{\i}}n, J.~C., Minniti, D., Gromadzki, M., et al.\ 2013, \aap, 557, L8
\bibitem[Beichman et al.(2014)]{Beichman2014} Beichman, C., Gelino, C.~R., Kirkpatrick, J.~D., et al.\ 2014, \apj, 783, 68
\bibitem[Bertin \& Arnouts(1996)]{Bertin1996} Bertin, E., \& Arnouts, S.\ 1996, \aaps, 117, 393 
\bibitem[Bertin et al.(2002)]{Bertin2002} Bertin, E., Mellier, Y., 
Radovich, M., et al.\ 2002, Astronomical Data Analysis Software and Systems 
XI, 281, 228 
\bibitem[Biller et al.(2013)]{Biller2013} Biller, B.~A., Crossfield, I.~J.~M., Mancini, L., et al. 2013, \apj, 778, L10
\bibitem[Bonnarel et al.(2000)]{Bonnarel2000} Bonnarel, F., Fernique, P., Bienaym{\'e}, O., et al.\ 2000, \aaps, 143, 33 
\bibitem[Boffin et al.(2014)]{Boffin2014} Boffin, H.~M.~J., Pourbaix, D., Mu{\v z}i{\'c}, K., et al.\ 2014, \aap, 561, L4
\bibitem[Burningham et al.(2011)]{Burningham2011} Burningham, B., Lucas, P.~W., Leggett, S.~K., et al.\ 2011, \mnras, 414, L90
\bibitem[Burgasser et al.(2006)]{Burgasser2006} Burgasser, A.~J., Geballe, T.~R., Leggett, S.~K., Kirkpatrick, J.~D., \& Golimowski, D.~A.\ 2006, \apj, 637, 1067 
\bibitem[Burgasser et al.(2014)]{Burgasser2014} Burgasser, A.~J., Gillon, M., Faherty, J.~K., et al.\ 2014, \apj, 785, 48  
\bibitem[Burrows et al.(2001)]{Burrows2001} Burrows, A., Hubbard, W.~B., Lunine, J.~I., \& Liebert, J.\ 2001, Reviews of Modern Physics, 73, 719 
\bibitem[Crossfield et al.(2014)]{Crossfield2014} Crossfield, I.~J.~M., Biller, B., Schlieder, J.~E., et al.\ 2014, \nat, 505, 654
\bibitem[Cushing et al.(2011)]{Cushing2011} Cushing, M.~C., Kirkpatrick, J.~D., Gelino, C.~R., et al.\ 2011, \apj, 743, 50 
\bibitem[Cushing et al.(2014)]{Cushing2014} Cushing, M.~C., Kirkpatrick, J.~D., Gelino, C.~R., et al.\ 2014, \aj, 147, 113 
\bibitem[de Bruijne(2014)]{Bruijne2014} de Bruijne, J.~H.~J.\ 2014, arXiv:1404.3896
\bibitem[Faherty et al.(2014a)]{Faherty2014} Faherty, J.~K., Beletsky, Y., Burgasser, A.~J., et al.\ 2014, \apj, 790, 90
\bibitem[Faherty et al.(2014b)]{Faherty2014b} Faherty, J.~K., Tinney, C.~G., Skemer, A., \& Monson, A.~J.\ 2014, \apjl, 793, L16
\bibitem[Kissler-Patig et al.(2008)]{Kissler2008} Kissler-Patig, M., Pirard, J.-F., Casali, M., et al.\ 2008, \aap, 491, 941 
\bibitem[Kirkpatrick et al.(1999)]{Kirkpatrick1999} Kirkpatrick, J.~D., Reid, I.~N., Liebert, J., et al.\ 1999, \apj, 519, 802 
\bibitem[Kirkpatrick et al.(2014)]{Kirkpatrick2014} Kirkpatrick, J.~D., Schneider, A., Fajardo-Acosta, S., et al.\ 2014, \apj, 783, 122
\bibitem[Kniazev et al.(2013)]{Kniazev2013} Kniazev, A.~Y., Vaisanen, P., Mu{\v z}i{\'c}, K., et al.\ 2013, \apj, 770, 124 
\bibitem[Kopytova et al.(2014)]{Kopytova2014} Kopytova, T., Crossfield, I., Deacon, N., et al.\ 2014, \apj, submitted
\bibitem[Leggett et al.(2013)]{Leggett2013} Leggett, S.~K., Morley, C.~V., Marley, M.~S., et al.\ 2013, \apj, 763, 130
\bibitem[Luhman(2012)]{Luhman2012} Luhman, K.~L.\ 2012, \araa, 50, 65
\bibitem[Luhman(2013)]{Luhman2013} Luhman, K.~L.\ 2013, \apjl, 767, L1  
\bibitem[Luhman(2014b)]{Luhman2014b} Luhman, K.~L.\ 2014, \apjl, 786, L18  
\bibitem[Luhman \& Sheppard(2014)]{Luhman2014c} Luhman, K.~L., \& Sheppard, S.~S.\ 2014, \apj, 787, 126 
\bibitem[Mart{\'{\i}}n et al.(1999)]{Martin1999} Mart{\'{\i}}n, E.~L., Delfosse, X., Basri, G., et al.\ 1999, \aj, 118, 2466 
\bibitem[Morales-Calder{\'o}n et al.(2006)]{Morales2006} Morales-Calder{\'o}n, M., Stauffer, J.~R., Kirkpatrick, J.~D., et al.\ 2006, \apj, 653, 1454 
\bibitem[Morley et al.(2014)]{Morley2014} Morley, C.~V., Marley, M.~S., Fortney, J.~J., Lupu, R., Saumon, D., Greene, T., \& Lodders, K. 2014, \apj, 787, 78
\bibitem[Nakajima et al.(1995)]{Nakajima1995} Nakajima, T., Oppenheimer, B.~R., Kulkarni, S.~R., et al.\ 1995, \nat, 378, 463 
\bibitem[Perez-Garrido et al.(2014)]{Perez2014} Perez-Garrido, A., Lodieu, N., Bejar, V.~J.~S., et al.\ 2014, arXiv:1405.5439
\bibitem[Pirard et al.(2004)]{Pirard2004} Pirard, J.-F., Kissler-Patig, M., Moorwood, A., et al.\ 2004, \procspie, 5492, 1763 
\bibitem[Radigan et al.(2014)]{Radigan2014} Radigan, J., Lafreni{\`e}re, D., Jayawardhana, R., \& Artigau, E.\ 2014, \apj, 793, 75
\bibitem[Rebolo et al.(1995)]{Rebolo1995} Rebolo, R., Zapatero Osorio, M.~R., \& Mart{\'{\i}}n, E.~L.\ 1995, \nat, 377, 129 
\bibitem[Scholz(2014)]{Scholz2014} Scholz, R.-D.\ 2014, \aap, 561, A113 
\bibitem[Smith et al.(2014)]{Smith2014} Smith, L., Lucas, P.~W., Bunce, R., et al.\ 2014, \mnras, 443, 2327
\bibitem[Stauffer et al.(1994)]{Stauffer1994} Stauffer, J.~R., Hamilton, D., \& Probst, R.~G.\ 1994, \aj, 108, 155
\bibitem[Torres et al.(2008)]{Torres2008} Torres, C.~A.~O., Quast, G.~R., Melo, C.~H.~F., \& Sterzik, M.~F.\ 2008, Handbook of Star Forming Regions, Volume II, 757
\bibitem[Werner et al.(2004)]{Werner2004} Werner, M.~W., Roellig, T.~L., \& Low, F.~J. et al. 2004, \apjs, 154, 1 
\bibitem[Wright et al.(2010)]{Wright2010} Wright, E.~L., Eisenhardt, P.~R.~M., \& Mainzer, A.~K. et al. 2010, \aj, 140, 1868 
\bibitem[Wright et al.(2014)]{Wright2014} Wright, E.~L., Mainzer, A., Kirkpatrick, J.~D., et al.\ 2014, \aj, 148, 82 
\end{thebibliography}
\end{document}